\documentclass[reprint,prd,nofootinbib,superscriptaddress]{revtex4}
\usepackage{CJK}
\usepackage{amsfonts}
\usepackage{amsmath,slashed}
\usepackage{amssymb}
\usepackage[english]{babel}
\usepackage{graphicx}
\usepackage{epsfig}
\usepackage{bm}
\usepackage{verbatim}
\usepackage[utf8]{inputenc}
\usepackage{booktabs}
\usepackage{multirow}
\usepackage{subfig}
\usepackage{slashed}
\usepackage{xcolor}
\usepackage[colorlinks=true,urlcolor=red,citecolor=red]{hyperref}
\newcommand{\hs}{\hspace*{0.3cm}}

\newcommand{\be}{\begin{equation}}
	\newcommand{\ee}{\end{equation}}
\newcommand{\bea}{\begin{eqnarray}}
	\newcommand{\eea}{\end{eqnarray}}
\newcommand{\ben}{\begin{enumerate}}
	\newcommand{\een}{\end{enumerate}}
\newcommand{\bit}{\begin{itemize}}
	\newcommand{\eit}{\end{itemize}}
\newcommand{\bde}{\begin{widetext}}
	\newcommand{\ede}{\end{widetext}}
\newcommand{\nn}{\nonumber}
\newcommand{\crn}{\nonumber \\}

\newcommand{\al}{\alpha}
\newcommand{\la}{\lambda}
\newcommand{\bet}{\beta}
\newcommand{\ga}{\gamma}
\newcommand{\va}{\varphi}

\newcommand{\pa}{\partial}

\newcommand{\fr}{\frac}
\newcommand{\sq}{\sqrt}
\newcommand{\bc}{\begin{center}}
	\newcommand{\ec}{\end{center}}

\newcommand{\ka}{\kappa}

\newcommand{\eq}{\eqref}
\newcommand{\lb}{\label}

\newcommand{\mathsym}[1]{{}}
\newcommand{\gev}{~\mathrm{GeV}}

\newcommand{\CP}{$C\kern-0.7pt P$}
\definecolor{bostonuniversityred}{rgb}{0.8, 0.0, 0.0}

\def\SM{$\mathrm{SU(3)_c \otimes SU(2)_L \otimes U(1)_Y}$ }

\def\gsim{\raise0.3ex\hbox{$\;>$\kern-0.75em\raise-1.1ex\hbox{$\sim\;$}}}
\def\lsim{\raise0.3ex\hbox{$\;<$\kern-0.75em\raise-1.1ex\hbox{$\sim\;$}}}

\def\SM{$\mathrm{SU(3)_c \otimes SU(2)_L \otimes U(1)_Y}$ }

\newcommand{\stai}{Subatomic Physics Research Group, Science and Technology Advanced Institute,\\
 Van Lang University, Ho Chi Minh City, Vietnam}
\newcommand{\steh}{Faculty of Applied Technology, Van Lang School of Technology,
\\
 Van Lang University, Ho Chi Minh City, Vietnam}

\begin{document}
		\title{Probing  axion in the DFSZ model}
\title{Probing  axion in the DFSZ model}
\author{H. N. Long
}
	\email{hoangngoclong@vlu.edu.vn 
	}
	\author{N. H. T. Nha
	}
	\email{nguyenhuathanhnha@vlu.edu.vn}	
\affiliation{\stai}
\affiliation{\steh}

	\date{\today }
	
\begin{abstract}
 We show that with the introduction of new right-handed neutrinos $N_R$, the Peccei--Quinn ($PQ$) transformation in the DFSZ (Dine--Fischler--Srednicki--Zhitnitsky) model depends only on the $PQ$ charge of the scalar singlet $\phi$, namely $PQ_\phi$, and is independent of the sine and cosine of $\arctan(v_u/v_d)$. The model consists of four new scalars: a charged Higgs boson $H^\pm$, a CP-odd scalar $A$, and two CP-even scalars with masses at the EW scale (of a few hundred GeV), and a superheavy inflaton $\Phi$.
 The scalar fields have been presented in the form of unitary matrices, in which there is no mixing between the axion and the Goldstone boson $G_Z$. As a result, there are two kinds of couplings between the axion and fermions, namely Yukawa-like interactions and anomaly-induced ones associated with derivative axion couplings.  {The model belongs to the DFSZ I kind since the
 axion - photon coupling with  $\fr E N = \fr 8 3$.} The trilinear Higgs self-coupling is investigated. We have shown that, in the parameter region where new scalar fields such as the singly charged Higgs boson $H^\pm$ and the CP-odd scalar $A$ have masses around 150 GeV and the vacuum expectation value of one Higgs doublet ($v_u$) is of the order of a few GeV, the coupling modifier $\kappa_\lambda$-a probe of new physics-is consistent with the current experimental constraints.

\end{abstract}
	
	
\maketitle
\noindent

\section{Introduction}
\label{Intro}
The $PQ$ mechanism is well-known as an elegant solution to eliminate the strong $CP$ puzzles. The simple extension of the $PQ$ mechanism is the DFSZ model, where the Standard Model (SM) is just extended by adding one doublet and one non-trivial singlet scalar \cite{DFSZ1,DFSZ2}. The	DFSZ models were proposed in the early 1980s when neutrinos were still thought of as massless. Consequently,  the model did not require right-handed neutrinos.
Nowadays, to provide masses to neutrinos, three right-handed neutrinos should be introduced, as in the models \cite{smash,ah} or the 3-3-1 model with an axion-like particle	\cite{alp331}. It is worth mentioning that the introduction of right-handed neutrinos provides not only neutrino issues but also solves the $g-2$ problem \cite{ah,alp331}.

There is one more advantage of the model, namely, the   PQ transformations depend on just the PQ charge of the scalar singlet $PQ_\phi$ without a combination of the latter and sinus/cosinus of $ \arctan (v_u/v_d)$ as in Refs. \cite{DFSZ1,DFSZ2,luzio,ah,bl1}.

In difference at Ref. \cite{ah}, in this work, each parts of the scalar sector, namely, the charged, CP-odd and CP-even scalars, are diagolanized by unitary matrices in which physical
fields are clearly orthogonal. This ensures that there is no mixing between the axion $a$ and the Goldstone boson $G_Z$ \cite{{giogi}}. In addition, this allows us to further study collider phenomenology.

The layout of the remainder of this work is as follows. In Section~\ref{Review}, we present the particle content, Yukawa interactions, and Higgs potential, which are the main ingredients for determining the PQ transformation. Section~\ref{scalar} is devoted to the determination of the scalar fields, with particular focus on the axion and the SM-like Higgs boson ($h$). In Section~\ref{PQchargeSec}, we show that, with the newly introduced right-handed neutrinos, the PQ transformations depend on only one parameter, $PQ_\phi$. The formula for the axion state is obtained by imposing the invariance under the PQ transformation. The axion couplings, such as anomaly-induced and Yukawa-like interactions, are presented in Section~\ref{axioncoup}. {Section~\ref{tri} is devoted to studying the Higgs self-coupling modifier of the model, and we summarize our results in the last section, Section~\ref{conc}.}

	
\section{The model}
	\label{Review}
	
	The model is based on the    {SM} gauge group  \SM
	with  fermion content   being three singlet heavy right-handed neutrinos as in \cite{DFSZ1,DFSZ2,ah}
	\bea
	Q_L & = & \left( \begin{array}{c}
		u_L\\
		d_L  \\
	\end{array} \right) \sim \left(3, 2, \fr 1 3\right)\,,\crn
	\psi_L & = & \left( \begin{array}{c}
		\nu_L\\
		l_L  \\
	\end{array} \right) \sim \left(1, 2, -1\right)\,,\label{j171}\\
	u_R & \sim & \left(3, 1, \fr 4 3\right)\,,\hs  d_R  \sim  \left(3, 1, -
	\fr 2 3\right) \,,\hs  l_R \sim  \left(1, 1, - 2\right)\,,
	\crn
	{N_R} & \sim & \left(1, 1, 0\right)\,.
	\eea
	Scalar sector   consists two doublets and one singlet as in the DFSZ model \cite{DFSZ1,DFSZ2}
	\bea
H_d & = & \left( \begin{array}{c}
		\al^+\\
		\al^0 \\
	\end{array} \right) \sim \left(1, 2, 1\right)\,,\label{j172}\\
	H_u & = & \left( \begin{array}{c}
		\bet^+\\
		\bet^0 \\
	\end{array} \right) \sim \left(1, 2, 1 \right)\,,\label{j172}\\
	\phi & \sim & \left(1, 1, 0\right)\,.
	\nn
	\eea
With the above scalar spectrum, the Lagrangian of scalar sector is determined as below \cite{DFSZ1,DFSZ2,ah}
	\bea
	\mathcal{L}_{scalar} & = & \fr 1 2  \pa_\mu\phi \pa^\mu\phi    {+ \sum_{i=u,d}\left(D_\mu H_i \right)^\dagger D^\mu H_i}
	-V(\phi,H_u,H_d)\,,
	\label{LagrangianScalars}
	\eea
where the second term helps to define the physical states of $W^\pm, Z$ bosons and photon while the last term is the scalar potential.
	
The electric charge operator is similar with itself in SM theory
\be
Q = T_3 + \fr Y 2 \,,
\label{chargeoperator}
\ee
with $T_3$ is generator of $SU(2)_L$ group and $Y$ is hyper-charge
	operator.

\subsection{Yukawa interactions and neutrino physics }\label{YukI}

Finally, the Yukawa interactions are
\bea 
\mathcal{L}_{Yukawa} & = &
	y_u^{H_u}\overline{Q}_L \widetilde{H}_u u_R+y_d^{H_d}\overline{Q}_L H_d d_R+y_l^{H_d}\overline{\Psi}_{L_l} H_d l_R
\crn
	&&+y_{N_{\alpha}}^l\overline{\Psi}_{L_l} \widetilde{H}_d N_{R_{\alpha}}
+ \left(y_N\right)_{a b} \overline{N}^C_{a R}N_{b R} \phi^*
+\text{H.c.} \,, 
	\lb{june143}
\eea

\begin{figure}
			[h!]
			\centering
			\begin{tabular}{c}
\includegraphics[width=9cm]{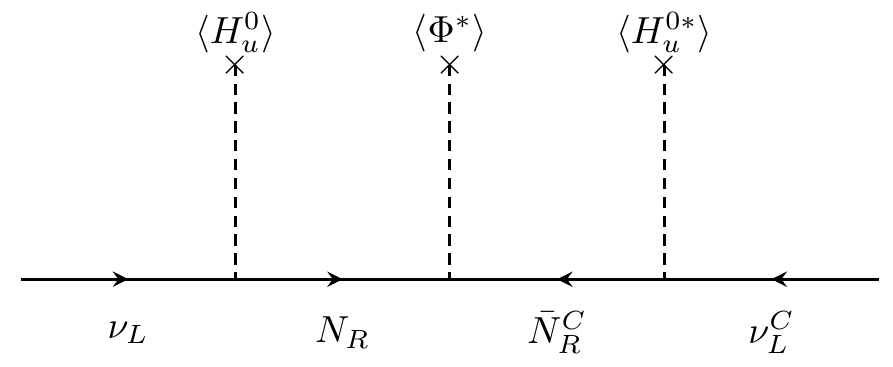}
			\end{tabular}%
\caption{Feynmann diagram illustrates type I seesaw mechanism that generates masses to light active neutrinos in the model}
	\label{type1ss}
\end{figure}

The $u$-type quarks get masses from $H_u$ while 
$d$-type quarks and charged leptons get masses from from $H_d$.
The neutrino masses are    generated by the below Lagrangian
\bea
- \mathcal{L}_{Yukawa}^{n} =  \left(y_\nu^D\right)_{a b}(\overline{\nu}_{a L} H_u^{0} N_{b R}
+ \overline{N}_{b R}H_u^{0 *} {\nu_{a L}}{^C})
+ \left(y_N\right)_{a b} (\overline{N}^C_{a R}N_{b R} \phi^* 
+\overline{N}_{b R}N^C_{a R} \phi)
\label{LYukNeutrino}
\eea
From  Eq.~\eq{LYukNeutrino}, it is shown that light active neutrino mass
matrices arisen from type I seesaw mechanism mediated by right-handed Majorana neutrinos (see figure \ref{type1ss}) has the form 		
\be
m_\nu = M_\nu^D M_N^{-1} (M_\nu^D)^T, \hs M_\nu^D = \fr{\langle H_u \rangle}{\sq{2}} y_\nu^D \,,\hs M_N = \sq{2}
\langle \phi \rangle  y_N\,.
\label{MassNeutrino}
\ee

In the case $M_R \gg M^D$, the diagonalization leads to \cite{ss1,ss2}
\be
m_{\nu} \simeq \fr{(M^D)^2}{M} \,, \hs M \simeq M_N \,.
\ee
As a result, one gets neutrino mass in the range 
$m_\nu \in \mathcal{O}(1)$ eV, when 
$M_N \sim 10^7 \gev$  and $M_D \sim \left(10^{-1} \div 10^{-2}\right) \gev $ \cite{ss3,alp331}. 

It is worth mentioning that the heavy neutrinos with mass around $10^7 \gev$ coupling to charged lepton and scalar [see Eq. \eq{YukDFSZ1}] produce  baryogenesis through leptogenesis. Therefore the derived model contains
very interesting features in neutrino physics.

\subsection{Higgs potential}
\label{Hpot}
		
The  last term in Eq.~(\ref{LagrangianScalars}) is scalar potential
 and given by  \cite{ah,eff}
\bea	
V\left(\phi, H_u, H_d\right) & = & \mu_d^2 H_d^\dag H_d + \mu_u^2 H_u^\dag H_u
+ \la_\phi \left(\phi^* \phi - \fr{V^2_\phi}{2}\right)^2 + \phi^* \phi
(k_d H_d^\dag H_d +  k_u H_u^\dag H_u)  \crn
&& + \la_d (H_d^\dag H_d)^2
+ \la_u (H_u^\dag H_u)^2 + \la [(H_d^\dag H_d )( H_u^\dag H_u) - (H_d^\dag H_u)( H_u^\dag H_d) ]\crn
&& - \left( \kappa \phi H_u^\dag  H_d + \text{H.c.} \right)
\label{ScalarPotential}
\eea

In order to analysis the scalar potential more easily, the scalars should be expanded around their the vacuum expectation values (VEVs) before substituting into Eq.~ 
\eq{ScalarPotential}. Hereafter, the following notations are used
\bea
H_d  & = & \left( \begin{array}{c}
\alpha^+\\
\fr{1}{\sqrt2}\left(v_d +R_D +i I_D \right)
\end{array} \right)   \,, \hs
H_u   =  \left( \begin{array}{c}
\beta^+\\
\fr{1}{\sqrt2}\left(v_u +R_U +i I_U \right)  \\
\end{array} \right)\,,
\label{ScalarDoublets}\\
\phi & = &  \fr{1}{\sqrt2}\left(v_\phi +R_\phi +i I_\phi \right)\,. \label{ScalarSinglet2}
\eea
where VEVs $ v_\phi \sim 10^{12}\, \gev$ \cite{DFSZ2,vphi1,vphi2,vphi3,vphi4},
 and $v_d, v_u$ are at electroweak (EW) scale.
The term associated with $\la$ consists  only {\it charged} scalar
couplings.

Within VEV structure in \eq{ScalarDoublets}, it follows that the masses of the SM gauge bosons are given by
\be
		m_W^2 = \frac{g^2}{4}\left(v_u^2 + v_d^2 \right)\,,\hs
		m_Z^2 = 
		\frac{m_W^2}{\cos^2 \theta_W}
		\,. \label{mZmWtt}
		\ee
and photon $A_\mu $ is  massless. Therefore
\be 
v_u^2 + v_d^2 = v^2 = 246^2 \, \gev^2\,.
\lb{jan211}
\ee

 The minimization of the Higgs potential yields
\bea
\la_\phi (v^2_\phi - V^2_\phi) + \fr{(k_d v^2_d + k_u v^2_u)}{ 2 }
- \fr{\kappa v_d v_u}{\sq{2} v_\phi} &=& 0 \,, \crn
\mu_u^2 + \fr{k_u}2  v^2_\phi  + \lambda_{u} v^2_u 
  - \fr{\kappa v_\phi v_d}{\sq{2} v_u}  &=& 0 \,,  \label{mcondition} \\
\mu_d^2 + \fr{k_d}2  v^2_\phi  + \lambda_{d} v^2_d 
  - \fr{\kappa v_\phi v_u}{\sq{2} v_d}  &=& 0 \,.
\nn
\eea

With  $\mu_i,  (i=u, d)$,  and $V_\phi $ that satisfy
 Eq.~\eq{mcondition} , are substituted into the scalar potential
  in Eq.~\eq{ScalarPotential} in order to determine the scalar mass
  squared matrices.

\section{Scalar fields}
\label{scalar}

\subsection{Charged scalars}\label{chargedscalars}
In the basis of $(\alpha^\pm, \beta^\pm)$, the mass squared matrix is
\begin{align}
M^2_{chagred} &= \left(
\begin{array}{cc}
 \frac{s_{\beta } \left(\lambda  c_{\beta } s_{\beta } v^2+\sqrt{2} \kappa  v_{\phi }\right)}{2 c_{\beta }} & -\frac{1}{2} \lambda  c_{\beta } s_{\beta }
   v^2-\frac{\kappa  v_{\phi }}{\sqrt{2}} \\
 -\frac{1}{2} \lambda  c_{\beta } s_{\beta } v^2-\frac{\kappa  v_{\phi }}{\sqrt{2}} & \frac{c_{\beta } \left(\lambda  c_{\beta } s_{\beta } v^2+\sqrt{2} \kappa 
   v_{\phi }\right)}{2 s_{\beta }} \\
\end{array}
\right).
\end{align}
Here  $t_{\beta}= \tan \bet = \fr{v_u}{v_d}$.
Consequently, the model consists of two mass eigenstates, corresponding to  a massless Goldstone boson $G^{\pm}_W$ of $W^{\pm}$ and a physical singly charged Higgs state $H^\pm$ with mass being determined as follows 
\be 
m^2_{H^\pm} = \frac{1}{2} \left(\lambda  v^2 +\frac{\sqrt{2} \kappa  v_{\phi }}{c_{\beta } s_{\beta }}\right)= \frac{v^2}{2}\left(\lambda   +\frac{\sqrt{2} \kappa  v_{\phi }}{v_u v_d}\right)\,.
\lb{dec271}
\ee
From \eq{dec271}, it follows
 \be \fr{\la} 2 + \fr{\sq{2}\kappa v_\phi}{ v^2 \sin 2 \bet } > 0\,.
 \lb{e161}
 \ee
 The relationship between original  and physical states  are shown below
\bea
\left(\begin{array}{c}
	\alpha^\pm \\ \beta^\pm
\end{array}\right) =\left(
\begin{array}{cc}
	c_{\beta } & -s_{\beta } \\
	s_{\beta } & c_{\beta } \\
\end{array}
\right)
\left(\begin{array}{c}
	G_W^\pm \\ H^\pm
\end{array}\right)\,. 
\lb{feb93}
\eea
\subsection{CP odd scalars}\label{CPoddscalars}
In the basis of $(I_{H_d^0},  I_{H_u^0},I_\phi)$, the respective 
squared mass matrix is

\bea
M_{odd}^2= 
 \fr{\kappa v_\phi v_u v_d}{\sq{2}} \left(
\begin{array}{ccc}
 \frac{1}{v^2_d} & - \frac{1}{v_d v_u}&\frac{1}{v_d v_\phi} \\
 - \fr{1}{v_d v_u} &  \frac{1}{v^2_u} & - \frac{1}{v_u v_\phi}\\
\fr{1}{v_d v_\phi} & -\frac{1}{v_u v_\phi} &  \frac{1}{v^2_\phi} \\
\end{array}
\right) \,.
\lb{dec272}
\eea

Defining that:
\begin{align}
\label{j151}
 \tan \theta = \fr{v_u v_d}{v_\phi v}
= \fr{\sin 2\bet \,  v}{2 v_\phi}
 \sim
\fr{v}{ v_\phi}\,.
\end{align}
then diagonalizing gives a transformation between the mass and original base as follows  


\bea
\left(\begin{array}{c}
	A \\ G_Z \\ a
			\end{array}\right) =C_a
			\left(\begin{array}{c}
	 I_{H_d^0}\\  I_{H_u^0} \\I_\phi 
			\end{array}\right)\, ,\; C_a= \left(
\begin{array}{ccc}
- c_{\theta } s_{\beta } & c_{\beta } c_{\theta } & s_{\theta } \\
 c_{\beta } & s_{\beta } & 0 \\
 - c_\bet s_\theta & s_\theta s_\bet & c_\theta
   \\
\end{array}
\right) \, \Rightarrow \left(\begin{array}{c}
 I_{H_d^0}\\  I_{H_u^0} \\I_\phi 
			\end{array}\right) =C^T_a
			\left(\begin{array}{c}
		A \\ G_Z \\ a
			\end{array}\right)
\label{dep}
\eea


Therefore
\be
a =  c_\theta  I_\phi  -  c_\bet  s_\theta   I_{H_d^0}+s_\bet    s_\theta  I_{H_u^0}
 \,.
\label{dep2}
\ee
Note that, as expected, $G_Z$ does not contain $I_\phi$.

For practical use, one has $(I_{H_d^0},\; I_{H_u^0} ,\;  I_\phi)^T =C_a (A, 	G_Z,\;  a,)^T$. Note here that the limit $\theta=0$ will give a consistent consequence that $I_\phi\equiv a$, namely the axion decouples from other Higgs bosons, similar to the case of the two-Higgs-doublet model (THDM).  Then, one has two massless scalars, namely one Goldstone boson $G_Z$ and an axion $a$, and one massive neutral CP-odd Higgs $A$, whose mass is given by
\bea m^2_A & = &\frac{\kappa  \left(c_{\beta }{}^2 s_{\beta }{}^2 v^2+v_{\phi }{}^2\right)}{\sqrt{2} c_{\beta } s_{\beta } v_{\phi }}\crn
&\simeq & \fr{\kappa  v_\phi v^2}{2 \sq{2} v_u v_d}=
\fr{\kappa  v_\phi}{ \sq{2} \sin 2\bet} 
\lb{dec282}
\eea
Hence to have $m_{A} = 130 \gev$, as  by recent experiments 
in \cite{A130},   it follows
\be 
\ka \simeq \fr{m^2_{A} \sq{2} \sin 2\bet  }{v_\phi} \leq \fr{m^2_{A} \sq{2}  }{v_\phi} \approx \fr{m^2_{A}}{v_\phi} \simeq 10^{-8} \gev = 10 \, \text{eV}\,.
\lb{sec112}
\ee
Note that in the frameworks of chiral perturbative Lagrangian  in the energy scale $ \lesssim 1 \, \gev$ \cite{chp1,chp2,chp3},
one gets axion mass squared 
at  leading order as follows \cite{weinberg,stern,luzio,luzio2,giogi,jan12}
\be  m_a^2 = \fr{m_u \, m_d}{(m_u + m_d)^2}\fr{m^2_\pi f^2_\pi}{f^2_a} \simeq 5.7 \left(\fr{10^{12} \, \gev}{f_a}  \right)
\mu \textrm{eV} \,.
\label{s158}
\ee

If one also considers contribution from $s$ quark, 
Eq. \eqref{s158} becomes \cite{sr}

\bea m_a &=& 4  \fr{f_\pi m_\pi}{f_a/N}\left[\fr{ m_u m_d m_s}{(m_u m_d + m_u m_s + m_d m_s)(m_u + m_d)}  \right]^{\fr 1 2}\crn
&\simeq & (1.2 \times 10^{-5} \, \textrm{eV})\left( \fr{10^{12} \, \gev}{f_a/N} \right)\,,
\label{s121}
\eea

Axion mass  is also discussed in \cite{amass1,amass2}.
\be
m_a = 5.70 \left[\fr{10^9 \, GeV}{f_a}\right] meV \, . 
\label{dec295} \ee


\subsection{CP even sector}

In the  basis of $ (R_{D}, R_{U}, R_\phi)$, the mass squared matrix is
\bea
M_{even}^2= \left(
\begin{array}{ccc}
 2 c_{\beta }{}^2 \lambda _d v^2+\frac{s_{\beta } v_{\phi } \kappa }{\sqrt{2} c_{\beta }} & -\frac{v_{\phi } \kappa }{\sqrt{2}} & c_{\beta } v_{\phi } \kappa _d
   v-\frac{s_{\beta } \kappa  v}{\sqrt{2}} \\
 -\frac{v_{\phi } \kappa }{\sqrt{2}} & 2 s_{\beta }{}^2 \lambda _u v^2+\frac{c_{\beta } v_{\phi } \kappa }{\sqrt{2} s_{\beta }} & s_{\beta } v_{\phi } \kappa _u
   v-\frac{c_{\beta } \kappa  v}{\sqrt{2}} \\
 c_{\beta } v_{\phi } \kappa _d v-\frac{s_{\beta } \kappa  v}{\sqrt{2}} & s_{\beta } v_{\phi } \kappa _u v-\frac{c_{\beta } \kappa  v}{\sqrt{2}} & 2 \lambda _{\phi }
   v_{\phi }{}^2+\frac{c_{\beta } s_{\beta } \kappa  v^2}{\sqrt{2} v_{\phi }} \\
\end{array}
\right)
\eea
Within limits $v_\phi \gg v\,, \ka k_d,, \ka k_u $, ($ \ka k_d,, \ka k_u \simeq \fr{m_A^2 v^2}{v^3_\phi} $) and following 
the procedure in Ref.  \cite{hn}), one obtains
the physical fields  given by
\bea
\left(\begin{array}{c}
	h \\ S \\  \Phi
			\end{array}\right) =\left( \begin{array}{ccc}
 c_{\al_2}	 &  c_{\al_1} s_{\al_2} & s_{\al_1} s_{\al_2}\\
- s_{\al_2} & c_{\al_1} c_{\al_2}& s_{\al_1} c_{\al_2}\\
 0  &- s_{\al_1} & c_{\al_1}
\end{array} \right)
			\left(\begin{array}{c}
	 R_{D}\\   R_{U}\\ R_\phi 
			\end{array}\right)\,,
\label{higgd}
\eea
where
\be 
 \tan 2 \al_1 = - \fr{k_d v_d}{\la_\phi v_\phi}\,,\hs
\tan 2 \al_2 = - \fr{c_{\al_1}\left(  \la  v_d  v_u   -  \fr{\kappa v_\phi }{ \sq{2}} \right) }{
c^2_{\al_1}  \la_d v^2_d -  \la_u v^2_u  + \fr{\kappa v_\phi }{2 \sq{2}}
\left(\tan \bet - \cot \bet\right)}\,.
\lb{jan134}
\ee

For practical use
\bea
\left(\begin{array}{c}
 R_{D}\\   R_{U}\\ R_\phi 
			\end{array}\right) =\left( \begin{array}{ccc}
 c_{\al_2}	 &  - s_{\al_2} & 0\\
  c_{\al_1} s_{\al_2} & c_{\al_1} c_{\al_2}&  - s_{\al_1}\\
s_{\al_1} s_{\al_2}   &s_{\al_1} c_{\al_2}   & c_{\al_1}
\end{array} \right)
			\left(\begin{array}{c}
	h \\ S \\  \Phi
			\end{array}\right)\,,
\label{higgd}
\eea
From \eq{dec282}, it follows

\bea m^2_A & = &\fr{\kappa  v_\phi v^2}{2 \sq{2} v_u v_d} \Rightarrow
\fr{\kappa  v_\phi}{ \sq{2}} = \fr{2 m^2_A v_u v_d}{v^2} 
\lb{jan131}
\eea
From \eq{dec271}, it follows
\be 
m^2_{H^\pm} = \frac{v^2}{2} \left(\lambda  +\frac{\sqrt{2} \kappa  v_{\phi }}{c_{\beta } s_{\beta v^2 }}\right) = \frac{v^2}{2} \left(\lambda 
 +\frac{\sqrt{2} \kappa  v_{\phi }}{v_u v_d}\right) =
 \frac{v^2}{2} \left(\lambda 
 +\frac{4 m^2_A }{v^2}\right)\,.
\lb{jan132}
\ee
Combination of  \eq{jan131} and \eq{jan132} yields 
\be 
\la = \fr 2{v^2} \left(   m^2_{H^\pm}  - 2  m^2_A \right)\,.
\lb{jan133}
\ee
Next, from \eq{jan134}, one has
\bea
\tan \al_2 & = & \fr{ \sin 2 \bet \left(   m^2_{H^\pm}  - 3  m^2_A
\right) }{
v^2 [  s^2_ \bet (\la_u  +  \la_d) -  \la_d] + m^2_A (1 - 2 s^2_ \bet)}\,,
 \lb{jan135}\\
\Rightarrow \sin \al_2 & =  & \fr{\sin 2 \bet \left(   m^2_{H^\pm}  - 3  m^2_A
\right)}{\left\{\sin^2 2 \bet \left(   m^2_{H^\pm}  - 3  m^2_A
\right)^2 + \left(v^2 [  s^2_ \bet (\la_u  +  \la_d) -  \la_d] + m^2_A (1 - 2 s^2_ \bet)\right)^2\right\}^{1/2}}
 \lb{a233}
\eea

The masses of physical fields are following
\bea 
m^2_h & = & 
 2  \la_d v^2 (1 - s^2_ \bet)  + 2 m^2_A s^2_ \bet
-   \fr{  \fr 1 2[\sin 2 \bet \left(   m^2_{H^\pm}  - 3  m^2_A
\right) ]^2 }{v^2 [  s^2_\bet (\la_u  +  \la_d) -  \la_d] + m^2_A (1 - 2 s^2_ \bet)} \,,
\label{jan181}\\
m^2_S & = & 2  \la_u v^2 s^2_ \bet  +  2 m^2_A (1- s^2_ \bet) 
+    \fr{  \fr 1 2[\sin 2 \bet \left(   m^2_{H^\pm}  - 3  m^2_A
\right) ]^2 }{v^2 [  s^2_\bet (\la_u  +  \la_d) -  \la_d] + m^2_A (1 - 2 s^2_ \bet)} \,,
\label{jan1182}\\
m^2_\Phi   & = &  \frac{\kappa  c_{\beta } s_{\beta } v^2}{\sqrt{2} v_{\phi }}+2 \lambda _{\phi } v_{\phi }{}^2 \approx 2 \la_\phi v^2_\phi\,. \label{jan1183}
\eea 

In the above expressions, Eqs   \eq{jan131}, \eq{jan132} and \eq{jan135} have been used.
Mass of inflaton $\Phi$ is: $m_\Phi = \sq{2 \la_\phi} v_\phi \, \approx 10^{11}\, \gev$ if  
$ \la_\phi = 10^{-2}$ as in Ref. \cite{ah}.

In the limit $v_\phi \gg v_d > v_u$, one has
\bea
H_d  & = & \left( \begin{array}{c}
G_{W^+}\\
\fr{1}{\sqrt2}\left(v_d + S +i G_Z \right)
\end{array} \right)   \,, \hs
H_u   =  \left( \begin{array}{c}
H^+\\
\fr{1}{\sqrt2}\left(v_u + h +i {A} \right)  \\
\end{array} \right)\,,
\lb{dec128}\\
\phi & = &  \fr{1}{\sqrt2}\left(v_\phi + \Phi +i a  \right)\,. \nn
\eea

\subsection{Mass spectrum of scalar sector}
\lb{mrel}
For future analysis, let us denote
\be \lambda_{d}\equiv t_\la \lambda_{u}\,.
\label{jan143}\ee
Then Eq. \eq{jan181} and Eq. \eq{jan1182} become
\bea 
m^2_h(\la_u,t_\la) & = &
 2   t_\la \la_u v^2 (1 - s^2_ \bet)  + 2 m^2_A s^2_\bet
-   \fr{  2 [  s_\bet \sq{1-s^2_ \bet} \left(   m^2_{H^\pm}  - 3 m^2_A 
\right) ]^2 }{ v^2   \la_u [s^2_\bet (t_\la + 1) -t_\la]+ m^2_A (1 - 2 s^2_ \bet)}
\label{apr291}\\
m^2_S(\la_u,t_\la) & = & 2  \la_u v^2 s^2_ \bet  +  2 m^2_A (1- s^2_ \bet) 
+     \fr{ 2 [  s_\bet \sq{1-s^2_ \bet} \left(   m^2_{H^\pm}  - 3 m^2_A 
\right) ]^2 }{ v^2   \la_u [s^2_\bet (t_\la + 1) -t_\la]+ m^2_A (1 - 2 s^2_ \bet)}\,.
\label{apr292}
\eea

It is interesting that from  \eq{apr291} and  \eq{apr292}, it follows
\be
m^2_S  =  2 m^2_A - m^2_h + 2 t_\la  \la_u v^2 + 2 v^2 \la_u(1 - 
t_\la) s^2_ \bet  \,.
\label{jan142}
\ee

Eq. \eq{jan142} shows that, if scalar self- interactions $\la_d$ and 
$\la_u$ are smaller the unity and mass of pseudoscalar $A$ being 
around few hundreds GeV, then
mass of new CP even scalar $S$ is in few hundreds GeV too. In our concrete case
below it is around 220 GeV.

\begin{figure}
			[h!]
			\centering
			\begin{tabular}{c}
\includegraphics[width=9cm]{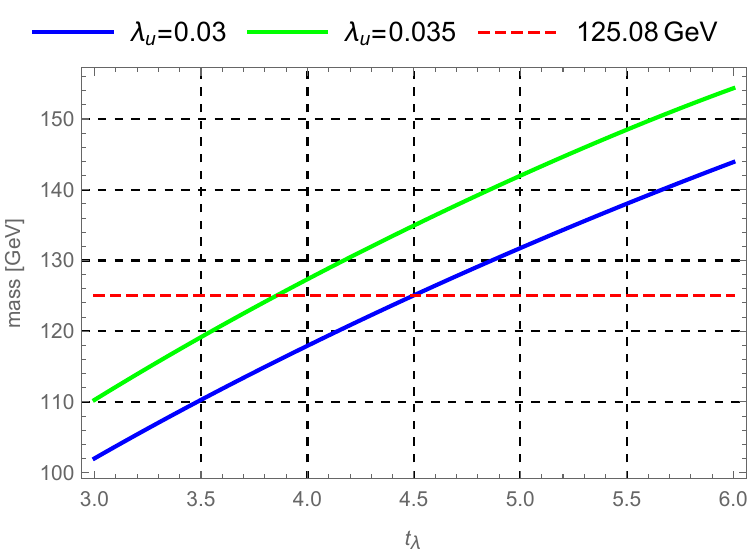}
			\end{tabular}%
\caption{ Mass  of SM-like Higgs  boson ($m_h$) as a function of $t_\la$, blue line for $\la_u=0.03$, green line for $\la_u=0.035$ and horizontal line is a
value $125.08 \gev$. As a consequence, for  $\la_u=0.03,\,  t_\la = 4.5$ and  
$\la_u=0.035,\,  t_\la = 3.9$}
	\label{mHigg}
\end{figure}

 In Fig. \ref{mHigg} we have plotted  mass  of SM-like Higgs boson ($m_h$) as a function of $t_\la$, for $\la_u = 0.03$ (blue line)and $\la_u = 0.035$ (green line)
 $m_A = 130 \, \gev\,,m_{H^\pm} = 155 \, \gev \,, \sin\bet = 0.0688$.  

The figure shows that  $\la_u=0.035,\,  t_\la = 3.9$ and  
$\la_u=0.03,\,  t_\la = 4.5$.

\begin{figure}
			[h!]
			\centering
			\begin{tabular}{cc}
\includegraphics[width=8.5cm]{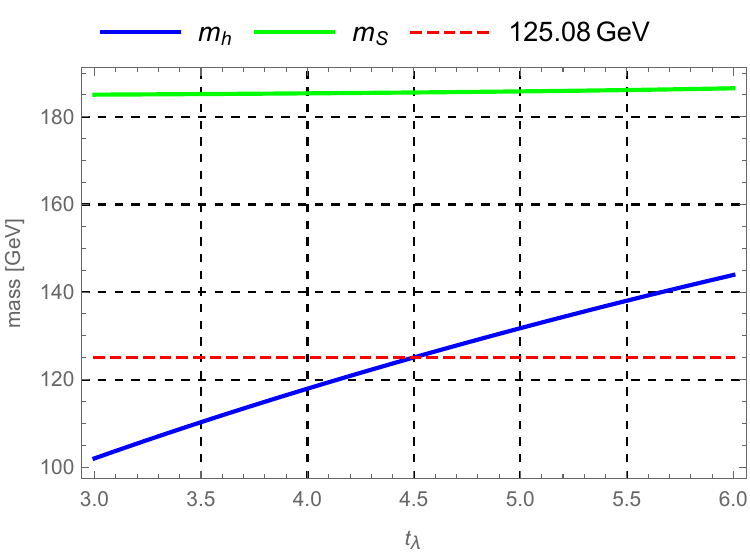} &\hspace{0.3cm}
\includegraphics[width=8.5cm]{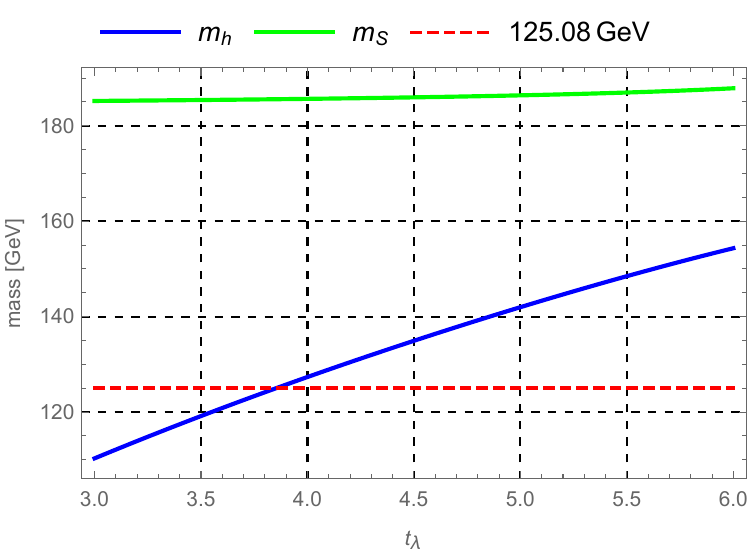}
			\end{tabular}%
\caption{Masses  of SM-like Higgs  boson ($m_h$)  and of  heavy scalar boson ($m_S$) are  as  functions of $\text{sin}_\beta$ at $t_\lambda $, blue line for SM-like Higgs boson, green line for $S$ Higgs boson.  The left panel is for  $\la_u = 0.03$,
and the right panel is  for  $\la_u = 0.035$}
	\label{mSHigg}
\end{figure}

In Fig. \ref{mSHigg}, we have plotted masses of SM-like Higgs boson ($m_h$)
and the heavy boson  ($m_S$) as  functions of $t_\la$ for $\la_u = 0.03$ in left panel and for $\la_u = 0.035$ in right
panel.

From  the result in Fig.\ref{mSHigg}, it follows that $m_S =  221.878 \, \gev$ for  $\la_u = 0.03$ and 
$m_S =  217.086 \, \gev$ for  $\la_u = 0.035$.

To finish this section, let us summarize 
\ben
\item In the CP-odd scalar sector: there are three bosons: one Goldstone boson 
 $G_Z$, 
  massless axion $a$ and one massive pseudoscalar $A$ with  mass
   around $\mathcal{O}(100) \gev$
 
\item In the scalar charged sector: two fields, one of which is a Goldstone boson
$G_W$ eaten by $W$ boson. Another field has mass at the EW scale

\item In CP-even scalar sector: one has one super heavy scalar $\Phi$ playing
 a role of inflaton with  mass around $10^{11}\, \gev$ 
 and new scalar $S$ with  mass around $220\, \gev$ and SM-like Higgs boson.

\item From formulas of charged and CP-odd scalars, it follows non-zero VEVs of scalar doublets. Taking masses of charged and CP-odd scalars to be 155 GeV and 130
GeV, respectively, one obtains $\la = - 0.323055$.
\een

Note that the bound of masses in this work is slightly higher than in
Ref. \cite{ma2hdm}.


\section{$PQ$ charges}\label{PQchargeSec}

From Yukawa couplings in \eq{june143} and \eq{LYukNeutrino}, one has
\bea \label{yuk}
- \mathcal{L}_{Yukawa} & = &
	y_u^{H_u}\overline{Q}_L \widetilde{H}_u u_R+y_d^{H_d}\overline{Q}_L H_d d_R+y_l^{H_d}\overline{\Psi}_{L_l} H_d l_R
\crn
	&&+y_{N_{\alpha}}^l\overline{\Psi}_{L_l} \widetilde{H}_d N_{R_{\alpha}}
+ \left(y_N\right)_{a b} \overline{N}^C_{a R}N_{b R} \phi^*
+\text{H.c.} \crn
&  =  &y_u^{H_u} ( \overline{u}_L H_u^{0 *} u_R  -
		\overline{d}_L H_u^- u_R + \overline{u}_R H_u^{0} u_L - \overline{u}_R H_u^{+} d_L ) \crn
&&+ y_d^{H_d} ( \overline{u}_L  H_d^{+} d_R +
		\overline{d}_L H_d^0 d_R + \overline{d}_R  H_d^{-} u_L + \overline{d}_R  H_d^{0*} d_L)\crn
&& +  y_l^{H_d}(\overline{\nu}_L H_d^{+} l_R
		+ \overline{l}_L H_d^{0} l_R + \overline{l}_R H_d^{-} \nu_L + \overline{l}_R H_d^{0*} l_L )\crn
&& +    { \left(y_\nu^N\right)_{a b}(\overline{\nu}_{a L} H_d^{0*}
			N_{b R} -  \overline{l}_{a L} H_d^{-} N_{b R}- \overline{N}_{b R} H_d^{+} l_{a L}
+ \overline{N}_{b R}H_d^{0 } \nu_{a L} )} \label{YukDFSZ1}\\
&&+     {\left(y_N\right)_{a b} (\overline{N}^C_{a R}N_{b R} \phi^* + (\overline{N}_{b R}N^C_{a R} \phi)}
\,.\nn
	\lb{dec133t}
\eea
Here we have denoted: $H_d^{\pm} \equiv \al^\pm $ 
and $ H_u^{\pm} \equiv \bet^\pm $.
In our work, we follow the rules in Ref.~
\cite{bl1} to define $PQ$ charges of all components in multiplets, as well as the $PQ$ charges of multiplets without the assumption that the $PQ$ charge of a multiplet is an average of the $PQ$ charges of the multiplet's components. This is because this assumption is automatically satisfied. Assuming that all the classical Lagrangian automatically possesses the global $U(1)_{PQ}$ symmetry and following the mentioned rules, we firstly require that Yukawa interactions must be preserved under a $U(1)_{PQ}$ transformation to get the relation below for the $PQ$ charge of a scalar and the $PQ$ charges of two fermions
\be
PQ_{ (\phi)} =   PQ_{(\bar{f}_L)} + PQ_{(f_R)}\,.
\label{PQscalar_fermions}
\ee

The invariance under PQ symmetry, the last term in \eq{ScalarPotential} yields
\be 
PQ_\phi  -  PQ_{H_u} + PQ_{H_d} = 0\,.
\label{jun141}
\ee

In summary, $PQ$ charges of particles in the model are shown in Table \ref{tab1}.
				\begin{table}[h!]
				\renewcommand{\arraystretch}{1.6}
					\resizebox{17cm}{!}{
						\begin{tabular}{|c|c|c|c|c|c|c|c|c|c|c|c|c|c|c|}
							\hline
& $Q_L$ & $\psi_L$ & $H_u$ & $H_d$ & $u_R$ & $d_R$ & $l_R$ & $\nu_L$ & $N_R$ & $H_u^0$ & ${\bet^+}$ & $H_d^0$ & ${\al^+}$ & $\phi$  \\ \hline							$U(1)_{PQ}$	& $-\frac{PQ_\phi}{4}$ &  $-\fr{PQ_\phi}{4}$ & $\fr{7 PQ_\phi}{4}$ 
&  $\fr{3 PQ_\phi}{4}$ &	$\fr{3 PQ_\phi}{2}$ & $ - PQ_\phi$ &$ - PQ_\phi$ &
$-\frac{3 PQ_\phi}{2}$ & $\frac{PQ_\phi}{2}$ & $ 3 PQ_\phi$ & $\fr{PQ_\phi}{2}$ & $ 2 PQ_\phi$ & $-\fr{PQ_\phi}{2}$ & $PQ_\phi$\\ \hline
\end{tabular}}
\caption{$U(1)_{PQ}$ charge assignments of the particle content of the model with only one parameter $PQ_\phi$ coming from the singlet scalar causing $U(1)_{PQ}$ spontaneous symmetry breaking.}			
\label{tab1}
\end{table}


\section{Axion couplings \label{axioncoup}}
\subsection{Anomaly couplings of axion}
In the model under consideration, the $QCD$ anomaly coefficient is defined as ~\cite{luzio}
\be
\label{nov4}
N=\sum_\mathcal{Q} N_\mathcal{Q} = \sum_\mathcal{Q} PQ({\mathcal{Q})}\,
 n_C (\mathcal{Q})\,  n_I(\mathcal{Q})\,  T(\cal{C}_{\mathcal{Q}}) \,,
\ee
in which,
	\begin{enumerate}
		\item $ n_C (\mathcal{Q})$ and $n_I(\mathcal{Q})$ denote the dimension of the  colour and weak isospin representations, respectively, for example:  ($n( {\bf  T) }= 3$ for triplet and $n ({\bf S})	= 1$ for singlet).
		\item $T(\cal{C}_\mathcal{Q})$ is the colour Dynkin index, for example
		$T(3)=1/2$.
	\end{enumerate}

Taking into account  $PQ q_L = - PQ q_R$ and values of $PQ$ charges in Tab.~\ref{tab1}, the $QCD$ anomaly coefficient is read as 
\bea
 N(DFSZ) & = & n_c     \fr 1 2 [ (PQ_{ Q_{ a L}}  - PQ_{ u_{a R})} +
 (PQ_{ Q_{ a L}}  - PQ_{ d_{a R})}] =  \fr 3 2 (- PQ_{ H_{u}} +  PQ_ {H_{d}}) = - \fr 3 2  PQ_\phi \,.
\label{N1}
\eea

Now we turn to  electromagnetic $[U(1)_Q]^2\times U(1)_{PQ}$ anomaly coefficient
\be
E  = \sum\limits_{i= charged}\left( PQ_{f_{iL}} - PQ_{f_{i R}}\right)(Q_{f_i})^2\,.
\label{EU1QU1PQ}
	\ee	
Taking $PQ$ charges from Table \ref{tab1} and the color number of quarks  $n_c $ to be three, one has
\bea
E_{DFSZ} & = &   n_c 3 [(PQ_{ u_{ a L}}  - PQ_{ u_{a R}}) ( \fr 2 3 )^2
+ (PQ_{ d_{ a L}}  - PQ_{ d_{a R}}) ( -\fr 1 3 )^2] +  3  (PQ_{ l_{ a L}}  - PQ_{ l_{a R}}) (-1)^2\crn
& = &   n_c 3 [ - PQ_{H_u} ( \fr 2 3 )^2
+ PQ_{H_d} ( -\fr 1 3 )^2] +  3  (PQ_{H_d} (-1)^2\crn
& = & - 4 PQ_{H_u} +  PQ_{H_d} + 3  PQ_{H_d} = 4 (- PQ_{H_u} +  PQ_{H_d}) =  - 4 PQ_{\phi}\,.
\label{Edfsz1}
\eea
Combination of  Eq.~\eq{N1} and Eq.~\eq{Edfsz1} yields  the ratio of these two above coefficients is a constant as
\be
		\frac{E_{DFSZ}}{N_{DFSZ}}  = {\frac{8}{3}}\,. \label{ENdfsz}
	\ee
	
The same as in the DFSZ I model \cite{luzio}.

Performing current algebra, one gets the axion mass given by \cite{sr}
\be
		g_{a\ga} =\fr{\al}{2 \pi f_a} \left(\fr E N - \fr{2}{3}\fr{4+\frac{m_u}{m_d} + \fr{m_u}{m_s} }{1+\fr{m_u}{m_d}  +\fr{m_u}{m_s}}\right) = \fr{\al}{2 \pi f_a} \left(\fr E N  - 1.92 \right) \,.
		\label{gagaDFSZ1}
	\ee

\subsection{Anomaly axion-fermion couplings}
		\label{yua2}
		In this work, we are interested in the anomaly couplings related to the $PQ$ transformation.  Let us start from the kinematic term of a fermion in the free Lagrangian
		
		
The $PQ$	transformations are as follows
		\bea
%
		\va & \rightarrow& \va^\prime = e^{ i \left(\fr{
		c_\va}{2 f_{pq}}\right) a} \,,\crn
		f_L &\rightarrow& f^\prime_L  =  e^{ - i  \left(\fr{c_f}{2 f_{pq}}\right) a} f_L\,,\hs
		{\bar f}_L \rightarrow {\bar f}^\prime_L  = {\bar f}_L  e^{  i  \left(\fr{c_f}{2 f_{pq}}\right) a}\,,\label{hay1}\\
		f_R &\rightarrow & f^\prime_R  =  e^{  i  \left(\fr{c_f}{2 f_{pq}}\right) a} f_R\,,\hs
		{\bar f}_R \rightarrow {\bar f}^\prime_R  = {\bar f}_R  e^{ - i  \left(\fr{c_f}{2 f_{pq}}\right) a}\,.  \nn
		\eea
		where $c_f$ and $f_{pq} \sim 10^{11} \gev$ are the $PQ$ charge of the fermion and the axion decay constant related to
		the scale of the $U(1)_{PQ}$ global symmetry breaking, respectively.

Thus the axion - fermion  derivative  couplings  have the form
\bea
\mathcal{L}_ {(f-a)} & = &  + \left(\fr{PQ_f}{ \, f_a}\right) \pa_\mu \, a \, \bar{f}\,   \,\ga^\mu 
 \ga_5 f \,,\hs f= Q_L, \psi_L, u_R, d_R, l_R, N_R \label{dec285}\\
& = &   \left(\fr{1}{ f_a}\right) \pa_\mu \, a \,\left( \bar{f}_L \, c_f \,   \,\ga^\mu 
  f_L + \bar{f}_R c_f \,   \,\ga^\mu 
  f_R \right)\,,
 \label{mar241}
\eea
where $c_f = V^\dag_{CKM} PQ_f V_{CKM}$
for quarks and $c_f = V^\dag_{PMNS} PQ_f V_{PMNS}$
for leptons.

Our result  coincides with that in Ref. \cite{azga}. 

\subsection{Yukawa-like axion-fermion couplings}
In Eq.~\eq{YukDFSZ1}, the interactions between axion and a pair of SM fermions at tree-level are arisen from Yukawa couplings containing neutral electroweak scalar.
\bea \label{yuk}
- \mathcal{L}_{Yukawa} & = &
	y_u^{H_u}\overline{Q}_L \widetilde{H}_u u_R+y_d^{H_d}\overline{Q}_L H_d d_R+y_l^{H_d}\overline{\Psi}_{L_l} H_d l_R
\crn
	&&+y_{N_{\alpha}}^l\overline{\Psi}_{L_l} \widetilde{H}_d N_{R_{\alpha}}
+ \left(y_N\right)_{a b} \overline{N}^C_{a R}N_{b R} \phi^*
+\text{H.c.} \,, 
	\lb{dec61}
\eea
with axion is included in the imaginary parts of neutral electrical scalars as shown in Eq.~\eq{dep}. The Yukawa coupling of axion with a pair of SM fermions is now rewritten as
\bea
-\mathcal{L}_{a\bar{f}f}^{Yuk} &=& i \frac{m_u}{v_u} I_{H_u^0} \bar{u} \gamma_5 u + i \frac{m_d}{v_d} I_{H_d^0} \bar{d} \gamma_5 d + i \frac{m_l}{v_d} I_{H_d^0} \bar{l} \gamma_5 l    {+ i\frac{m_\nu}{v_u} I_{H_u^0} \bar{\nu} \gamma_5 N_{R}}+     { i \fr{m_N}{v_\phi}
I_\phi   \overline{N}^C_{ R} \ga_5 N_R}+ H.c\crn
& = &
  i\, a \sin \theta \left(\frac{m_u}{v_u} \,s_\bet  \bar{u} \gamma_5 u - 
 \frac{m_d}{v_d} \, c_\bet \bar{d} \gamma_5 d - \frac{m_l}{v_d} c_\bet  \bar{l} \gamma_5 l   + \cot \theta 
  \fr{m_N}{v_\phi}    \overline{N}_R^C \ga_5 N_R \right) \equiv  i c_{af}
  a \,   \bar{f} \gamma_5 f   .
\label{yukff}
\eea

Note that no Yukawa coupling of top-up quarks (u,c,t) with axion and
$ c_{af} \sim  \fr 1{v_\phi}$.

 Interaction Lagrangian as follows
\bea
\mathcal{L}_{int} & = & \fr{g^2}2  W^+_\mu W^{ - \mu}\{H^\pm H^\mp + G_W^\pm  G_W^\mp
+\fr 1 2 [ 2 v_u(c_{\al_2}	h   - s_{\al_2} S) + (c_{\al_2}	h   - s_{\al_2} S)^2
+ ( s_{\beta } G_Z  -c_{\beta } c_{\theta } A )^2\crn
& + &  2 v_d ( c_{\al_1} s_{\al_2} h + c_{\al_1} c_{\al_2} S  - s_{\al_1}
\Phi) + ( c_{\al_1} s_{\al_2} h + c_{\al_1} c_{\al_2} S  - s_{\al_1}
\Phi)^2 + ( c_{\beta } G_Z  -s_{\theta } a + c_{\theta } s_{\beta } A)^2] \}
\lb{feb43}\\
& + &  \fr{g^2}4  [(1 - 2 s^2_w)\fr{ Z_\mu}{c_W} + 2 s_W A_\mu]^2 
 (H^\pm H^\mp + G_W^\pm  G_W^\mp)] \lb{feb44t}\\
 & + &  \fr{g^2}{8 c^2_W} Z^\mu Z_\mu[ 2 v_u(c_{\al_2}	h   - s_{\al_2} S) + (c_{\al_2}	h   - s_{\al_2} S)^2
+ ( s_{\beta } G_Z  -c_{\beta } c_{\theta } A )^2\crn
& + &  2 v_d ( c_{\al_1} s_{\al_2} h + c_{\al_1} c_{\al_2} S  - s_{\al_1}
\Phi) + ( c_{\al_1} s_{\al_2} h + c_{\al_1} c_{\al_2} S  - s_{\al_1}
\Phi)^2 + ( c_{\beta } G_Z  -s_{\theta } a + c_{\theta } s_{\beta } A)^2] 
\lb{feb45t}
\eea

From \eq{feb44t} and  \eq{feb45t}, one sees that no coupling between axion and photon.

\section{Trilinear  Higgs self-couplings as a probe of new physics}
\label{tri}
The Higgs sector plays a crucial role in Particle Physics as a source for
masses of gauge bosons and fermions.  Note that the origin of gauge boson
 masses through spontaneous symmetry breaking (SSB) is also an automatic 
solution of the theory's renormalizability. One of the problems for the SM is unpredictable CP violation, and this leads to the THDM proposed half a century ago.

At present, the scalar sector with its self-couplings is a prime target to search 
for new physics. The trilinear self-coupling $\la_{hhh}$ is one of the flagship measurements at ATLAS and CMS  at CERN ~\cite{3h1}. For this aim, the coupling modifier defined as the ratio between the above-mentioned couplings in the SM and its extension is determined as follows
 \be \ka_\la\equiv \fr{\la_{hhh}}{\la_{hhh}^{\text{SM}0}}\,.
\lb{a1}
\ee 
Here $\lambda_{hhh}^{\text{SM} 0}$ is the tree-level SM prediction. 

The present bound for this parameter is given as ~\cite{3h1}

 \be -0.71<\ka_\la < 6.1 \hs\textrm{at}\, \hs  95\% \, C.L\ \,.
 \lb{j221}
 \ee 
The value will be a  target in the future linear colliders ~\cite{3h2,3h3}.

The trilinear Higgs self-coupling, $\lambda_{hhh}$, is extensively studied in frameworks of the THDM ~\cite{3hES1,3hES2,3h5}.

For further presentation, let us briefly mention on the SM potential
\bea V_{SM}(\Phi) & = & -\mu^2 \Phi^\dag \Phi + \la (\Phi^\dag \Phi)^2
\lb{j222}
 \eea 
where $ \Phi = \left( \phi^+, \fr 1{\sq{2}}(v + R_\phi + i I_\phi)\right)^T$.

After symmetry breaking, one gets
\bea
V_{SM}(\phi) & =  &- 
 \la v^2 h^2  + \la v h^3 + \cdots
 =   \fr 1 2 m_h^2 \, h^2 + \fr{ m_h^2}{2 v} h^3  +\cdots\,.
\lb{j223}
\eea
where $m_h^2 = 2 \la v^2$,

From ~\eq{j223}, it follows the Higgs self-coupling in the SM is
\be \lambda_{hhh}^{\text{SM}}= - \fr{3 m_h^2}v\,.
\lb{j224}
\ee

Substitution of Eq. ~\eq{higgd} into potential in ~\eq{ScalarPotential} yields trililear coupling
in the model under consideration as follows

\bea
 \la^{DFSZ}(hhh) & = & - 3\{c^3_{\al 2} s_{\al 1} v_\phi k_d +  c^3_{\al 2} c^2_{\al 1} 
s_{\al 1}v_\phi k_u  + 2 c^3_{\al 2}  s^3_{\al 1} v_\phi \la_\phi \crn
&+& c^3_{\al 2} c_\bet s^2_{\al 1} k_d v  + 2 c^3_{\al 2} c_\bet  \la_d v  +  2 c^3_{\al 1}   c^3_{\al 2}  s^2_{\al 1} s_\bet \la_u v  lb{j251}\\
&+& c_{\al 1}  c^2_{\al 2}  s_{\al 1} [- \sq{2} c_{\al 2} \ka + 
c_{\al 2}s_{\al 1} s_\bet  k_u v  ] \}\,.\nn
\eea 
Taking into account that $k_d = k_u = \fr{v^2}{v^2_\phi}$ ~\cite{ah} and $\sin \al_1 \propto \fr{v}{v_\phi}$, one obtains
\be
 \la^{DFSZ}(hhh)  =  -    6  c^3_{\al 2}  c_\bet  \la_d v  + \mathcal{O}
 \left( \fr{v^2}{v^2_\phi} \right)\,.
\lb{a2}
\ee

From \eq{jan135}, it follows
\bea
 c_{\al_2} = \fr{v^2 [  s^2_ \bet (\la_u  +  \la_d) -  \la_d] + m^2_A (1 - 2 s^2_ \bet)}{\{[ v^2 (  s^2_ \bet (\la_u  +  \la_d)
  -  \la_d) + m^2_A (1 - 2 s^2_ \bet) ]^2 +[ \sin 2 \bet \left(   m^2_{H^\pm}  - 3  m^2_A
\right)]^2 \}^{\fr 1 2} }\,.
 \lb{j302}
 \eea
Substituting of \eq{j302} into \eq{a2} ones gets finally
\bea   \la^{DFSZ}(hhh) & = &  -6 c_\bet  \la_d \, v \left(\fr{v^2 [  s^2_ \bet (\la_u  +  \la_d) -  \la_d] + m^2_A (1 - 2 s^2_ \bet)}{\{[ v^2 (  s^2_ \bet (\la_u  +  \la_d)
  -  \la_d) + m^2_A (1 - 2 s^2_ \bet) ]^2 +[ \sin 2 \bet \left(   m^2_{H^\pm}  - 3  m^2_A
\right)]^2 \}^{\fr 1 2} } \right)^3   \,.
 \lb{j253}   
 \eea
Substitution of Eqs. ~\eq{j224}, \eq{j253} into \eq{a1} yields
\bea  
\ka^{DFSZ}_\la  & = &-\fr{2 v  \la^{DFSZ}(hhh)}{ m_h^2}\crn
& = & \fr{12    c_\bet  \la_d \, v^2} {m_h^2}
\left(\fr{v^2 [  s^2_ \bet (\la_u  +  \la_d) -  \la_d] + m^2_A (1 - 2 s^2_ \bet)}{\{[ v^2 (  s^2_ \bet (\la_u  +  \la_d)
  -  \la_d) + m^2_A (1 - 2 s^2_ \bet) ]^2 +[ \sin 2 \bet \left(   m^2_{H^\pm}  - 3  m^2_A
\right)]^2 \}^{\fr 1 2} } \right)^3 \,.
\lb{j258}
\eea

Looking at formula \eq{j258}, we realize that the coupling modifier $\ka^{DFSZ}_\la$
depends on five parameter, namely: $\sin \beta, \la_d, \la_u, m_{H^\pm} $ and
$m_A$. We remind that the above parameters also appeared in the expressions of massive $CP$-even scalars $h$ and $S$. We have also taken into account the result
in Ref. ~\cite{ah} to assume the mass of the charged scalar $m_{H^\pm} = 155 \, \gev$.
From the expression of the square mass of the SM Higgs boson in \eq{apr291}, we have
 found  that $\sin \beta \approx 0.0688$, which means $v_u $ is in the range of few GeV.

\begin{figure}
			[h!]
			\centering
			\begin{tabular}{c}
\includegraphics[width=9cm]{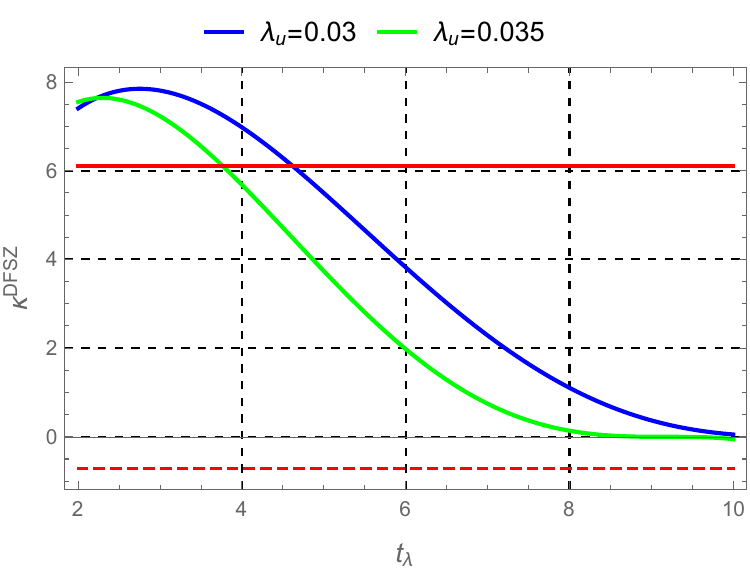}
			\end{tabular}%
\caption{ $\ka^{DFSZ}_\la $ as a function of $t_\la$, where $m_A=130 \, \gev, m_{H^\pm}= 155 \, \gev \, ,  \sin \bet = 0.0688$, $\la_u = 0.03\, \text{(blue line), and}\, \la_u = 0.035$ (green line). Two horizontal lines in red, thick and dashed, are experimental limits appearing in Eq. \eq{j221}.}
	\label{Fig4}
\end{figure}

In Fig. ~\ref{Fig4}   we have plotted $\ka^{DFSZ}_\la$ as a function of $t_\la$

Fig. \ref{Fig4} shows that for $\la=0.03$, the value $t_\la \geq 3.8$, while for
$\la=0.035$, the value $t_\la \geq 4.9$. The new physics arizes in the region 
$t_\la \in 4 \div 6$.

\begin{figure}
			[h!]
			\centering
			\begin{tabular}{cc}
\includegraphics[width=8.5cm]{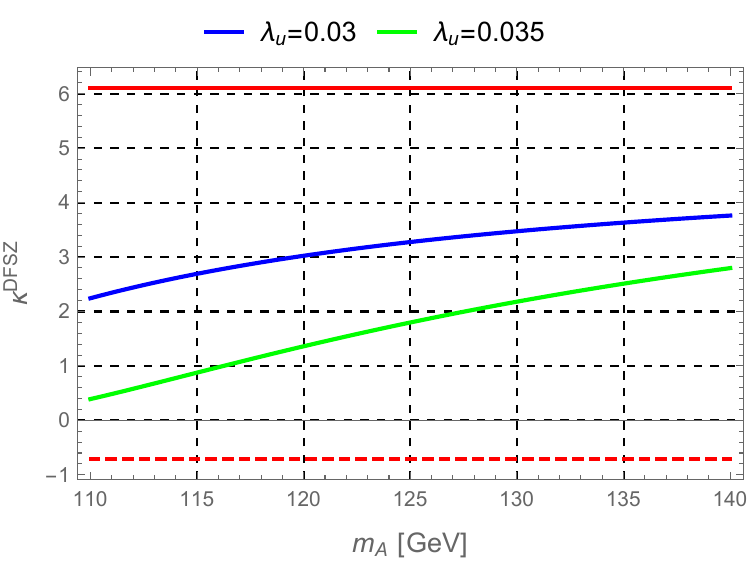} &\hspace{0.3cm}

\includegraphics[width=8.5cm]{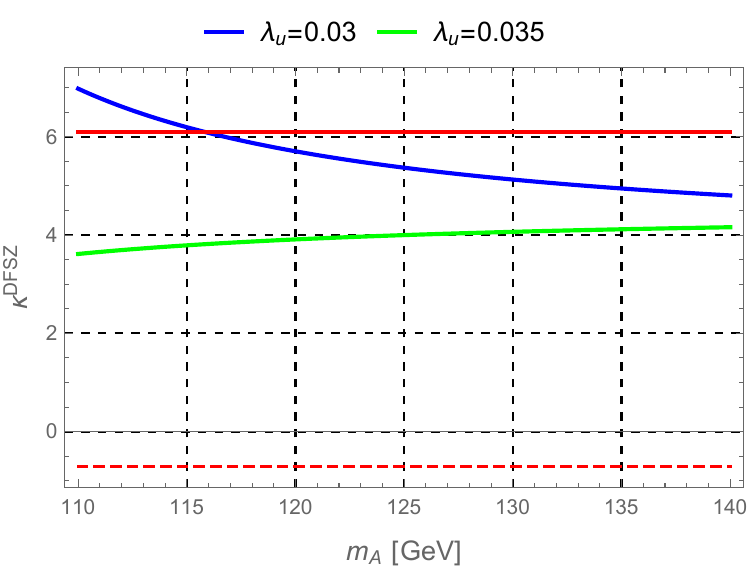}
			\end{tabular}%
\caption{ $\ka^{DFSZ}_\la $ as a function of $m_A$, where 
$ m_{H^\pm}= 155\, \gev, \, t_\la = 4.9$ (left-panel) and $t_\la = 3.9$ (right-panel),\,  $\sin \bet = 0.0688$, $\la_u = 0.03$ (blue line), and $\la_u = 0.035$ (green line). Two horizontal lines in red, thick and dashed, are experimental limits appearing in Eq. \eq{j221}.}
	\label{Fig5}
\end{figure}

In Fig. ~\ref{Fig5}   we have plotted $\ka^{DFSZ}_\la$ as a function of mass of  pseudoscalar  boson $m_A$

From Fig. ~\ref{Fig5}, it follows that, for $t_\la = 4.9$ the experimental up-limit is satisfied for
mass of CP-odd scalar $A$ leaving in the range from 110 GeV to 140 GeV, while for
 $t_\la = 3.9$, the situation is different for $\la_u=0.03$ and $\la_u=0.036$.
 
\section{Conclusions and Outlooks\label{conc}}
In this study, we have focused on the axion feature in the DFSZ  model with new right-handed neutrinos $N_R$. The  PQ transformation depends only on the PQ charge of the scalar singlet $\phi$.  The Yukawa interactions of the model lead to its axion-photon coupling being the same as in  the DFSZ I model (with $\fr E N = \fr 8 3$).
 The model consists of four new scalars: charged $H^\pm$, 
  and CP-even with mass around a few hundred GeV,
   and a super heavy inflaton $\Phi$.  The scalar fields have been presented in the form of unitary matrices,  in which there is no mixing among the axion and Goldstone boson $G_Z$.  As a result, there are two kinds of couplings between axion and fermions, namely Yukawa-like interactions and anomaly ones associated with the derivative of axion. 

 In this work, assuming  $m_A = 130 \, \gev\,,m_{H^\pm} = 155 \, \gev $ and
 $ \sin_\bet = 0.0688$  (i.e. $v_u \approx 3 \, \gev$), we get values of $\la_u  $  around $0.03$ and $t_\la$ around $4.5$. However, there is a special case in which the denominator of the last term in the expression of $m^2_h$ in \eq{jan181} vanishes, which should be considered in detail. 
 
The coupling modifier $\ka_\la$ - the parameter for searching New Physics has been 
investigated, and we have shown that the parameter in the model under consideration is well satisfied with the recent experimental limit of ATLAS and CMS.

As final comment we mention that the model can solve the following problems 
\ben
\item Neutrino mass and mixing
\item The muon $(g-2)$ anomaly
\item Inflation in Cosmology
\item Higgs physics
\een
Hence the model is devoted for future study.

\section*{Acknowledgements}
The authors thank Dr. L.T. Hue for his positive comments.
This research has received funding from
National Foundation for Science and Technology Development (NAFOSTED) under grant number 103.01-2025.04.


\begin{thebibliography}{99}
\bibitem{DFSZ2} M. Dine, W. Fischler, M. Srednicki, Phys. Lett. B {\bf 104} (1981) 199.
						
\bibitem{DFSZ1} A. Zhitnitsky, Sov. J. Nucl. Phys. {\bf 31} (1980) 260.

\bibitem{smash} G. Ballesteros, J. Redondo,   A.  Ringwald, and  C. Tamarit,  
JCAP 08 (2017) 001 , arXiv: 1610.01639 [hep-ph]

\bibitem{ah} M. Ahmadvand, F. Hajkarim,
Eur. Phys. J. C (2023) 83: 1021, arXiv:2302.09610 [hep-ph]


\bibitem{alp331} V. H. Binh, D. T. Binh,  A. E. C\'arcamo Hern\'andez, D. T. Huong, D. V. Soa, and  H. N. Long, Phys. Rev. D {\bf 107},  095030 (2023),  arXiv:2007.05004[hep-ph].

\bibitem{luzio} L. Di Luzio, M.  Giannotti, E. Nardi and L. Visinelli,  Phys. Rept. 870 (2020) 1,  arXiv:: 2003.01100 [hep-ph]

\bibitem{bl1} V. H. Binh and H. N. Long, 
Theoretical and Mathematical Physics (TMF), vol. 228, No 1, 121-139,  https://doi.org/10.4213/tmf11126, 
arXiv:2511.01781.

\bibitem{giogi} H. Georgi, D. B. Kaplan and L.  Randall,  Phys. Lett. \textbf{B 169} (1986) 73.

\bibitem{ss1} M. Gell-Mann, P. Ramond, and R. Slansky, Conf. Proc. C790927, 315 (1979), 1306.4669.

\bibitem{ss2} R. N. Mohapatra and G. Senjanovic, Phys. Rev. Lett. 44, 912 (1980).			

\bibitem{ss3}J. G. Ferreira, C. A de S. Pires, J. G. Rodrigues, P. S. Rodrigues da Silva,  Phys. Lett. B 771 (2017) 199,   arXiv: 1612.01463 [hep-ph]
		
\bibitem{eff} P. Ciafaloni, D. Espriu,
The Effective Lagrangian of the Two Higgs Doublet Model, Phys.Rev. D56 (1997) 1752-1760,  arXiv:hep-ph/9612383
doi 10.1103/PhysRevD.56.1752


\bibitem{vphi1} J. Preskill, M. B. Wise and F.Wilczek, Phys. Lett. B120, 127 (1983);
						
\bibitem{vphi2} L. F. Abbott and P. Sikivie, Phys. Lett. B120, 133 (1983);
						
\bibitem{vphi3} M. Dine and W. Fischler, Phys. Lett. B120, 137 (1983);
						
\bibitem{vphi4} G. Raffelt and D. Seckel, Phys. Rev. Lett. 60, 1793 (1988).

\bibitem{A130} The CMS collaboration, Search for an exotic decay of the Higgs boson to a
pair of light pseudoscalars in the final state of two
muons and two $\tau$ leptons in proton-proton collisions
at $\sq{s} = 13$ TeV , JHEP 11 (2018) 018,  ePrint: 1805.04865.
						
						
						
						

						

\bibitem{chp1} H. Sazdjian,  Introduction to chiral symmetry in QCD, EPJ Web Conf. 137 (2017) 02001, arXiv1 612.04078 [hep-ph].

\bibitem{chp2}  S. Scherer, Introduction to Chiral Perturbation Theory,  Adv. Nucl. Phys. 27 (2003) 277, 
arXiv:hep-ph/0210398  

\bibitem{chp3} L.  Di Luzio, G. Martinelli, G.  Piazza, Phys. Rev. Lett. 126, 241801 (2021), arXiv:2101.10330 [hep-ph].

\bibitem{stern} J. Stern, R. Zaoui, Regge pole and scattering lengths, Nucl. Phys. B 17 (1970) 253 - 266.


						

\bibitem{luzio2} L. Di Luzio and G. Piazza, JHEP 12 (2022) 041,  arXiv:2206. 04061.



\bibitem{jan12} Y. Giraldo, R. Martinez, E. Rojas, Juan C. Salazar, Eur. Phys. J. C  82 (2022) 1131,  
arXiv:2007.05653.

\bibitem{weinberg} S. Weinberg, Phys. Rev. Lett. 40 (1978) 223-226.

 \bibitem{sr} M. Srednicki, Nucl. Phys. B 260 (1985) 689.

 \bibitem{amass1} M. Gorghetto, G. Villadoro, 
 JHEP 03 (2019) 033,  arXiv:1812.01008 .

 \bibitem{amass2} G. Grilli di Cortona,  E. Hard,  J. P.  Vega and G. Villadoro, JHEP 01 (2016) 034, arXiv: 1511.02867.


\bibitem{hn} Le Tho Hue, Le Duc Ninh, 
Mod. Phys. Lett. A, Vol. 31, No. 10 (2016) 1650062,	arXiv:1510.00302.




\bibitem{ma2hdm} Chih-Ting Lu, K. Cheung, D. Kim, S. Lee, J. Song, Can a pseudoscalar with a mass of 365 GeV in the 2HDM explain the CMS  excess? arXiv:2601.12806 

	\bibitem{azga} A. Biekötter, K. Mimasu, Axions and Axion-like particles: collider searches, 
	arXiv:2508.19358.


\bibitem{cms} The CMS Collaboration, Measurement of the Higgs boson total decay width using the $H \to  WW \to e\nu \mu \nu$ decay channel in proton-proton collisions at $ \sq{s} = 13$ TeV, Phys. Rev. D 113 (2026) 092014, arXiv:2601.05168 [hep-ex]
 
\bibitem{pdg} P. A. Zyla et al, Prog. Theor. Exp. Phys. 2020, 083C01 (2020).


\bibitem{3h1} G. Aad, et al, Combination of ATLAS and CMS searches for Higgs
 boson pair production at $\sq{s} =$ TeV, CERN-EP-2026-011,
arXiv:2602.23991 [hep-ex]
 \bibitem{3h2}  G. Aad, et al, Highlights of the HL-LHC physics projections by ATLAS and CMS, arXiv:2504.00672
 \bibitem{3h3}  H. Abramowicz, A Linear Collider Vision for the Future of Particle Physics,(2025), arXiv:2503.19983
 





\bibitem{3hES1} S. Kanemura, S. Kiyoura, Y. Okada, E. Senaha, C.-P. Yuan,
 Phys. Lett. B558 (2003) 157,
arXiv:hep-ph/0211308.
 
\bibitem{3hES2} S. Kanemura, Y. Okada, E. Senaha, C.-P. Yuan,
Phys. Rev. D70 (2004) 115002,
arXiv:0408364.


\bibitem{3h5} J. Braathen, F. Egle, A. V. Schaeidt,
 Eur. Phys. J. Plus (2026) 141:688, 
 arXiv:2604.13922 

\end{thebibliography}
\end{document}